\RequirePackage{fixltx2e}
\documentclass[a4paper]{isp}
\usepackage[american]{babel}
\usepackage{graphicx}
\usepackage{amsmath, amsthm}
\usepackage{url}
\usepackage[capitalise,nameinlink]{cleveref}
\usepackage{xcolor}

\def\apj{ApJ\,  }
\def\apjs{ApJS  }

\def\mnras{MNRAS\,  }

\def\pasp{PASP  }
\def\pasa{PASA  }

\def\rmxaa{Revista Mexicana de Astronomia y Astrofisica}

\def\sun{\hbox{$\odot$}}
\def\sunmass{\mathcal{M}_{\sun}}

\begin{document}
\pdfgentounicode=1
\title
{
The   truncated Lindley distribution
with applications in astrophysics
}

\author{L. Zaninetti}
\institute{
Physics Department,
 via P.Giuria 1, I-10125 Turin,Italy \\
 \email{zaninetti@ph.unito.it}
}

\maketitle

\begin {abstract}
This paper reviews  the Lindley distribution
and then introduces the scale and the double
truncation.
The unknown parameters of the
truncated Lindley distribution are evaluated
with the
maximum likelihood
estimators.
An application  of the Lindley distribution
with scale is done to the initial  mass function for stars.
The magnitude version of the Lindley distribution with scale is
applied  to the luminosity function for the Sloan Digital Sky Survey (SDSS)
 galaxies
and to the photometric maximum of the 2MASS Redshift Survey 
(2MRS) galaxies.
The truncated Lindley  luminosity  function  allows us
to model the Malquist  bias  of the  2MRS galaxies.
\end{abstract}

\section{Introduction}

The Lindley distribution is defined by one parameter and
was introduced
to study  the difference  between fiducial distribution
and posterior distribution, see  \cite{Lindley1958,Lindley1965}.
Its  detailed properties
such as moments, cumulants, characteristic function, failure rate function
and $\dots$
can be found in \cite{Ghitany2008}.
We now briefly  outline some new trends, among others,
for this distribution.
A three parameter generalization of the Lindley distribution
has  been analyzed by \cite{Zakerzadeh2009},
the truncated versions of the Lindley distribution
has  been studied by \cite{Singh2014}, and
the estimation of  the parameters of the generalized
Lindley distribution has been done by \cite{Gui2016}
and
a three-parameter Lindley distribution
has been introduced by \cite{Shanker2017}.
A careful analysis  of these applications  in
the various fields of the natural sciences
has revealed that
the Lindley distribution has not
yet been applied to astrophysics.
Usually the mathematicians introduce many parameters,
which  characterize  statistical
distributions.
In contrast, applications  in the real world
require  fewer parameters,  such as  mean value
and variance.
The rapid development of computers has
allowed to simulate  the statistical distributions
through  the generation of  random numbers, but this
requires  the evaluation of the inverse  of
the distribution function.
A first  example  of an astrophysical application
for a probability density function (PDF)
is represented by the initial mass function for the stars (IMF).
The distribution in mass of the stars
has  been fitted with a power law.
This started with \cite{Salpeter1955},
who suggested that
$\xi ( {{m}}) \propto   {{m}}^{-\alpha}$
where $\xi   ( {{m}})$ represents  the probability
of having  a mass between $ {{m}}$ and
$ {{m}}+d{{m}}$;
He found  $\alpha= 2.35$
in the range $10  {M}_{\sun}~>~ {M} \geq  1  {M}_{\sun}$.
Subsequent research has started to analyze the initial  mass
function (IMF) with three  power laws, see
\cite{Scalo1986,Kroupa1993,Binney1998}, and four power laws, see
\cite{Kroupa2001}.
The  approach to the IMF using a continuous
distribution has been modeled by the lognormal distribution in
order to fit both the range of the stars and  the brown dwarfs
(BDs) regime, see \cite{Chabrier2003},
by the beta distribution, see \cite{Zaninetti2013a},
by the truncated gamma distribution, see \cite{Zaninetti2013e}
and
by the truncated lognormal distribution, see \cite{Zaninetti2017d}.
The previous  analysis raises  the
following questions:
\begin{itemize}
\item
Is it possible to find  the constant  of normalization
for  a left  and right truncated Lindley PDF?
\item
Is it possible to derive an analytical  expression
for the mean of
a left  and right truncated Lindley PDF
\item
Is  a left  and right truncated Lindley PDF a model
for the IMF  and for a   sample  of masses?
\end{itemize}
A second   example  of an astrophysical application
for a PDF is given
by the luminosity function (LF) for galaxies.
The Schechter function  was the first LF for galaxies
to be introduced, see
\cite{schechter}.
Over the years,
other LFs for galaxies  have been suggested, such as a
two-component Schechter-like LF,
see~\cite{Driver1996},
the hybrid Schechter+power-law LF
to fit the faint end of the K-band, see~\cite{Bell2003},
and the  double  Schechter LF,
see~\cite{Blanton_2005}.
To  improve the flexibility
at the bright  end, a new parameter
$\eta$ was  introduced in the Schechter LF,
see \cite{Alcaniz2004}.
A third astrophysical application is in the
photometric maximum  visible in the
number of cluster of galaxies as function of the redshift;
for example,
see Figure 7 in \cite{Coil2011}
where the number of galaxies as function of the redshift
is plotted
and
Figure 2 in \cite{Jimeno2015}
where the number of clusters for three catalogs are reported
as function of the redshift.
The theoretical explanation of this effect is
the joint distribution in redshift and
and flux  for galaxies;
see formula (5.133) in
 \cite{Peebles1993} or
 formula (1.104) in
 \cite{pad}
or
 formula (1.117)
in
\cite{Padmanabhan_III_2002}.
Despite this  theoretical background, the  photometric maximum
has been poorly analyzed.
A fourth astrophysical application is in
the range in absolute magnitude of galaxies versus the redshift
visible in the various catalogs;
for example, see Figure 9 in \cite{Coil2011}.
The mass of the stars in the IMF,
the luminosity
of galaxy in the LF
and the absolute magnitude of galaxy  in a given range of redshift
vary between a minimum and a maximum value.
This discussion suggests the introduction
of finite boundaries for the Lindley  IMF and LF
rather than the usual zero and infinity
following a  pattern similar to the introduction of
a left truncated beta LF; see \cite{Zaninetti2014d},
and for
a  left and right  truncated Schechter LF
luminosity function, see \cite{Zaninetti2017a}.

This paper  reviews
the original  Lindley distribution
in Section \ref{lindleysec}. It
introduces
 the scaling in Section \ref{lindleyscalesec}
and
the  double   truncation
in Section \ref{lindleytruncatedsec}.
The applications to the  astrophysics are
developed for  the IMF,
see Section \ref{lindleyimfsec},
and for the luminosity function (LF) for galaxies,
see Section \ref{lindleylfsec}.

\section{The Lindley family}

We present a
family of distributions of gradually increasing
complexity.

\subsection{Lindley distribution}
\label{lindleysec}
Let $X$ be a random variable
defined in
$[0, \infty]$;
the {\em Lindley}
probability density function (PDF), $f(x)$,
is
\begin{equation}
f (x;c) =
\frac
{
{c}^{2}{{\rm e}^{-cx}} \left( x+1 \right)
}
{
1+c
}
\quad ,
\end{equation}
the  distribution function (DF), $F(x)$,
 is
\begin{equation}
F (x;c) =
1- \left( 1+{\frac {cx}{1+c}} \right) {{\rm e}^{-cx}}
\quad,
\end{equation}
where $c>0$.
At $x=0$ $f(0) = {\frac {{c}^{2}}{1+c}}$ and not zero.

The average  value or mean, $\mu$,  is
\begin{equation}
\mu (c)=
\frac
{
2+c
}
{
c \left( 1+c \right)
}
\quad ,
\end{equation}
the variance, $\sigma^2$, is
\begin{equation}
\sigma^2(c)=
\frac
{
{c}^{2}+4\,c+2
}
{
{c}^{2} \left( 1+c \right) ^{2}
}
\quad .
\end{equation}
The rth moment about the origin
and  an approximation of the median are
reported in Appendix \ref{appendixdl}.

\begin{figure*}
\begin{center}
\includegraphics[width=10cm,angle=-90]{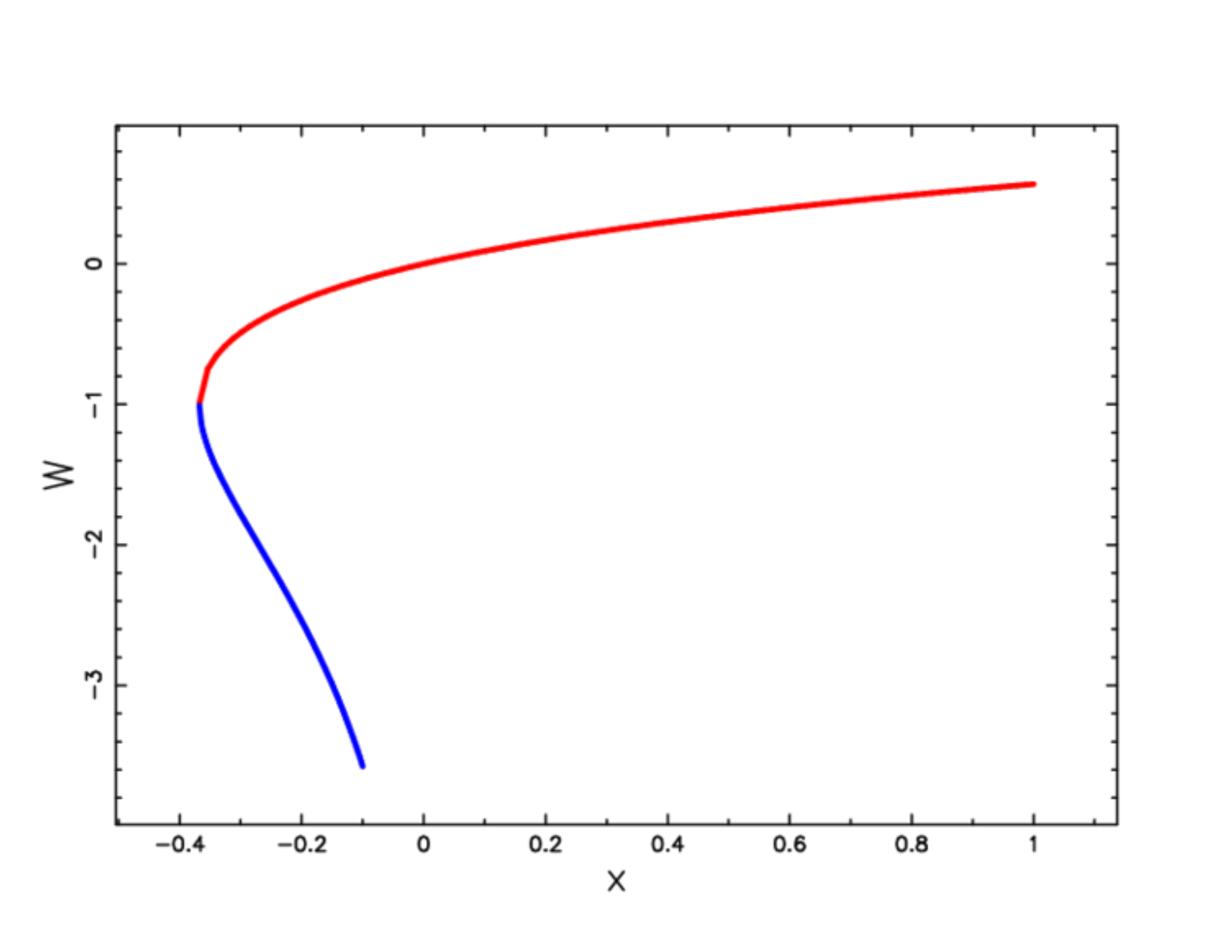}
\end{center}
\caption
{
Wp(x) (red line) and Wm(x) (blue line).
}
\label{lambertw_function}
\end{figure*}

The random generation of the Lindley variate X:c
is given by
\begin{equation}
X:c \approx
-{\frac {{\rm W} \left( \left( {\it R}-1 \right)  \left( 1+c
 \right) {{\rm e}^{-1-c}}\right)+1+c}{c}}
\quad ,
\label{xrandom}
\end{equation}
where $W$
is  the Lambert W function, after~\cite{Lambert_1758},
and R the unit rectangular variate R.
The Lambert W function according to \cite{NIST2010}
is defined as
\begin{equation}
We^{W}=x
\quad.
\end{equation}
The principal  branch, $Wp(x)$,  and
the other branch,
$Wm (x)$,
 of the Lambert W-function
can evaluated
with the Halley method
\begin{equation}
w_{n+1} =
w_{{n}}-
\frac
{
(w_{{n}}{{\rm e}^{w_{{n}}}}-x)
}
{
\left(  \left( w_{{n}}+1
 \right) {{\rm e}^{w_{{n}}}}-{\frac { \left( w_{{n}}+2 \right)
 \left( w_{{n}}{{\rm e}^{w_{{n}}}}-x \right) }{2\,w_{{n}}+2}} \right)
}
\quad ,
\end{equation}
see   \cite{Corless1996,Corless1997,Dence2013}.
The two branches of the Lambert W-function
are reported in Figure \ref{lambertw_function}.

\begin{figure*}
\begin{center}
\includegraphics[width=10cm,angle=-90]{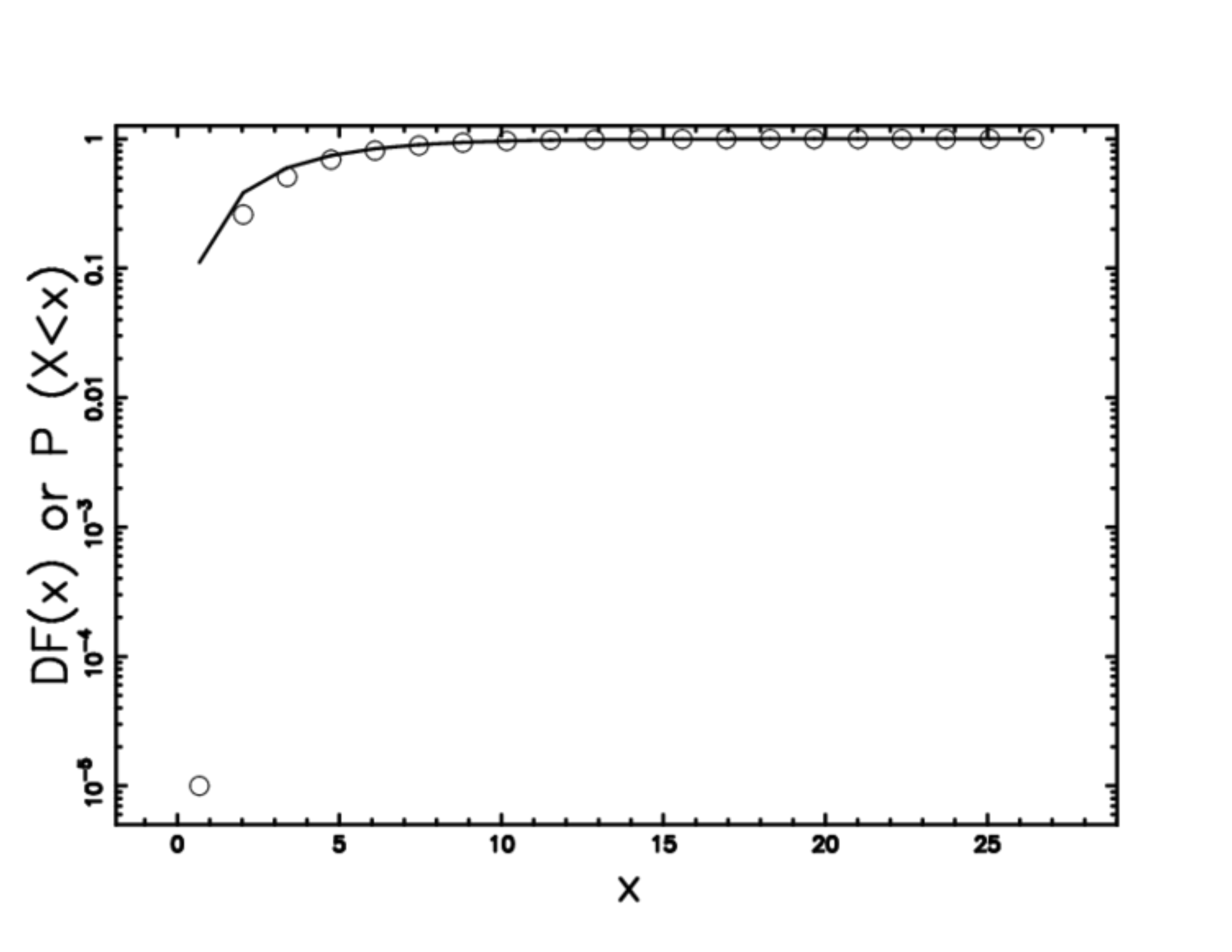}
\end{center}
\caption
{
Histogram of the simulated Lindley PDF
generated according to formula (\ref{xrandom})
 and
theoretical Lindley PDF (full line), 100000 random points and $c$=0.5.
}
\label{lindley_simulation}
\end{figure*}

A typical  simulation of the Lindley PDF is
reported in Figure \ref{lindley_simulation}.

The experimental sample  consists of the data $x_i$ with
$i$  varying between  1 and  $n$;
the sample mean, $\bar{x}$,
is
\begin{equation}
\bar{x} =\frac{1}{n} \sum_{i=1}^{n} x_i
\quad ,
\label{xmsample}
\end{equation}
the unbiased sample variance, $s^2$, is
\begin{equation}
s^2 = \frac{1}{n-1}  \sum_{i=1}^{n} (x_i - \bar{x})^2
\quad ,
\label{variancesample}
\end{equation}
and the sample $r$th moment  about the origin, $\bar{x}_r$,
is
\begin{equation}
\bar{x}_r = \frac{1}{n} \sum_{i=1}^{n} (x_i)^r
\quad .
\label{rmoment}
\end{equation}
The parameter $c$ can be obtained by the following
match
\begin{equation}
\mu_1  =\bar{x}_1
\quad ,
\end{equation}
and therefore
\begin{equation}
\widehat{c}  =
\frac
{
-\mu_{{1}}+1+\sqrt {{\mu_{{1}}}^{2}+6\,\mu_{{1}}+1}
}
{
2\,\mu_{{1}}
}
\quad .
\end{equation}

\subsection{The Lindley distribution with scale}

\label{lindleyscalesec}
We now introduce the scale $b$ in the Lindley distribution
and the PDF, $f_s(x;b,c)$,
is
\begin{equation}
f_s (x;b,c) =
\frac
{
{c}^{2}{{\rm e}^{-{\frac {cx}{b}}}} \left( x+b \right)
}
{
{b}^{2} \left( 1+c \right)
}
\quad ,
\label{lindleypdfscaling}
\end{equation}
the DF, $F_s(x;b,c)$, is
\begin{equation}
F_s (x;b,c) =
\frac
{
-{{\rm e}^{-{\frac {cx}{b}}}}bc-{{\rm e}^{-{\frac {cx}{b}}}}cx-{
{\rm e}^{-{\frac {cx}{b}}}}b+cb+b
}
{
b \left( 1+c \right)
}
\quad .
\end{equation}
The mean, $\mu_s(c,b)$,  is
\begin{equation}
\mu_s (c,b)=
{\frac {b \left( c+2 \right) }{c \left( 1+c \right) }}
\quad ,
\end{equation}
and the variance, $\sigma_s^2(c,b)$, is
\begin{equation}
\sigma^2_s(c,b)=
{\frac {{b}^{2} \left( {c}^{2}+4\,c+2 \right) }{{c}^{2} \left( 1+c
 \right) ^{2}}}
\quad .
\end{equation}
At $x=0$ $f_s(0) = {\frac {{c}^{2}}{ \left( 1+c \right) b}}$.

The rth moment about the origin
is reported in Appendix \ref{appendixldscale}.

The parameters $b$ and  $c$ can be obtained by the following
match
\begin{subequations}
\begin{align}
\mu_1       & = \bar{x}_1  \\
\sigma^2    & = s^2 \quad ,
\end{align}
\end{subequations}
which means
\begin{equation}
\widehat{c}
=
{\frac {-3\,{{\it \bar{x}}}^{3}+3\,{\it \bar{x}}\,{\it s^2}+\sqrt {2}\sqrt {
 \left( {{\it \bar{x}}}^{2}-{\it s^2} \right)  \left( 2\,{{\it \bar{x}}}^{2}-
{\it s^2} \right) ^{2}}}{{{\it \bar{x}}}^{2}-{\it s^2}}}
\quad ,
\end{equation}
and
\begin{equation}
\widehat{b}
=
\frac{1}{2}\,{\frac {1}{{\it \bar{x}}\, \left( {{\it \bar{x}}}^{2}-{\it s^2} \right)
} \left( \sqrt {2}\sqrt { \left( {{\it \bar{x}}}^{2}-{\it s^2} \right)
 \left( 2\,{{\it \bar{x}}}^{2}-{\it s^2} \right) ^{2}}+ \left( {{\it \bar{x}}
}^{2}-{\it s^2} \right)  \left( \sqrt {2}\sqrt {{\frac {{{\it s^2}}^{2}
}{{{\it \bar{x}}}^{2}-{\it s^2}}}}-4\,{\it \bar{x}} \right)  \right) }
\quad .
\end{equation}
The inequality  $s^2\,>  {\bar{x}}^2/2 $
makes both $\widehat{b}$  and  $\widehat{c}$ negatives
and, therefore, the sample is not  suitable
for a fit with Lindley distribution with scale.

\subsection{The  truncated Lindley distribution with scale}

\label{lindleytruncatedsec}
Let $X$ be a random variable
defined in
$[x_l,x_u]$;
the {\em truncated }
PDF, $f_t(x;b,c,x_{l},x_{u})$,
see   \cite{Singh2014,Hassanein2016},
is
\begin{equation}
f_t(x;b,c,x_{l},x_{u})=
{\frac {f_{{s}} \left( x \right) }{F_{{s}} \left( x_{{u}} \right) -F_{
{s}} \left( x_{{l}} \right) }}
\quad ,
\label{pdflindleytruncated}
\end{equation}
and the DF, $F_t(x;b,c,x_{l},x_{u})$,
\begin{equation}
F_t(x;b,c,x_{l},x_{u})=
{\frac {F_{{s}} \left( x \right)-F_{{s}} \left( x_l \right) }{F_{{s}} \left( x_{{u}} \right) -F_{
{s}} \left( x_{{l}} \right) }}
\quad .
\end{equation}
The inequality which fixes the range of existence
is $\infty>x_u>x>x_l>0$.

The first moment about the origin,
$\mu^{\prime}_{1,t}(b,c,x_l,x_u)$,
is
\begin{equation}
\mu^{\prime}_{1,t}(b,c,x_l,x_u) =
\frac
{
MN1
}
{
c \left( cb{{\rm e}^{{\frac {cx_{{l}}}{b}}}}+cx_{{u}}{{\rm e}^{{\frac
{cx_{{l}}}{b}}}}-cb{{\rm e}^{{\frac {cx_{{u}}}{b}}}}-cx_{{l}}{{\rm e}^
{{\frac {cx_{{u}}}{b}}}}+b{{\rm e}^{{\frac {cx_{{l}}}{b}}}}-b{{\rm e}^
{{\frac {cx_{{u}}}{b}}}} \right)
}
\quad ,
\end{equation}
where
\begin{eqnarray}
MN1=
{{\rm e}^{{\frac {cx_{{l}}}{b}}}}b{c}^{2}x_{{u}}+{{\rm e}^{{\frac {cx_
{{l}}}{b}}}}{c}^{2}{x_{{u}}}^{2}-{{\rm e}^{{\frac {cx_{{u}}}{b}}}}b{c}
^{2}x_{{l}}-{{\rm e}^{{\frac {cx_{{u}}}{b}}}}{c}^{2}{x_{{l}}}^{2}+{
{\rm e}^{{\frac {cx_{{l}}}{b}}}}{b}^{2}c+2\,{{\rm e}^{{\frac {cx_{{l}}
}{b}}}}bcx_{{u}}-{{\rm e}^{{\frac {cx_{{u}}}{b}}}}{b}^{2}c
\nonumber \\
-2\,{{\rm e}
^{{\frac {cx_{{u}}}{b}}}}bcx_{{l}}+2\,{{\rm e}^{{\frac {cx_{{l}}}{b}}}
}{b}^{2}-2\,{{\rm e}^{{\frac {cx_{{u}}}{b}}}}{b}^{2}
\quad ,
\end{eqnarray}
and the second
 moment about the origin,
$\mu^{\prime}_{2,t}(b,c,x_l,x_u)$,
is
\begin{equation}
 \mu^{\prime}_{2,t}(b,c,x_l,x_u) =
\frac
{
MN2
}
{
{c}^{2} \left( cb{{\rm e}^{{\frac {cx_{{l}}}{b}}}}+cx_{{u}}{{\rm e}^{{
\frac {cx_{{l}}}{b}}}}-cb{{\rm e}^{{\frac {cx_{{u}}}{b}}}}-cx_{{l}}{
{\rm e}^{{\frac {cx_{{u}}}{b}}}}+b{{\rm e}^{{\frac {cx_{{l}}}{b}}}}-b{
{\rm e}^{{\frac {cx_{{u}}}{b}}}} \right)
}
\quad ,
\end{equation}

\begin{eqnarray}
MN2=
{{\rm e}^{{\frac {cx_{{l}}}{b}}}}b{c}^{3}{x_{{u}}}^{2}+{{\rm e}^{{
\frac {cx_{{l}}}{b}}}}{c}^{3}{x_{{u}}}^{3}-{{\rm e}^{{\frac {cx_{{u}}
}{b}}}}b{c}^{3}{x_{{l}}}^{2}-{{\rm e}^{{\frac {cx_{{u}}}{b}}}}{c}^{3}{
x_{{l}}}^{3}+2\,{{\rm e}^{{\frac {cx_{{l}}}{b}}}}{b}^{2}{c}^{2}x_{{u}}
\nonumber \\
+3\,{{\rm e}^{{\frac {cx_{{l}}}{b}}}}b{c}^{2}{x_{{u}}}^{2}-2\,{{\rm e}
^{{\frac {cx_{{u}}}{b}}}}{b}^{2}{c}^{2}x_{{l}}-3\,{{\rm e}^{{\frac {cx
_{{u}}}{b}}}}b{c}^{2}{x_{{l}}}^{2}+2\,{{\rm e}^{{\frac {cx_{{l}}}{b}}}
}{b}^{3}c+6\,{{\rm e}^{{\frac {cx_{{l}}}{b}}}}{b}^{2}cx_{{u}}
\nonumber \\
-2\,{
{\rm e}^{{\frac {cx_{{u}}}{b}}}}{b}^{3}c-6\,{{\rm e}^{{\frac {cx_{{u}}
}{b}}}}{b}^{2}cx_{{l}}+6\,{{\rm e}^{{\frac {cx_{{l}}}{b}}}}{b}^{3}-6\,
{{\rm e}^{{\frac {cx_{{u}}}{b}}}}{b}^{3}
\quad .
\end{eqnarray}
The  variance, $\sigma^2_t(b,c,x_l,x_u)$,
is evaluated as
\begin{equation}
\sigma^2_t(b,c,x_l,x_u) =
 \mu^{\prime}_{2,t} -(\mu^{\prime}_{1,t})^2
\quad .
\end{equation}
The parameters $b$ and $c$ can be evaluated with the
 maximum likelihood estimators (MLE), see
Section~\ref{appendixmle}.

\section{The IMF for stars}

\label{lindleyimfsec}

The IMF for stars is  actually fitted with three
and four power laws, see
\cite{Hopkins2018,Hosek2019}.
The piece-wise broken inverse power law IMF
is
\begin{equation}
p(m) \propto m^{-\alpha_i} \quad,
\end {equation}
each zone  being  characterized  by a different
exponent  ${\alpha_i}$
and two boundaries $m_i$ and $m_{i+1}$.
To have a PDF
normalized  to unity, one must have
\begin{equation}
\sum _{i=1,n}  \int_{m_i}^{m_{i+1}} c_i m^{-\alpha_i} dm =1
\quad.
\label{uno}
\end{equation}
The number of parameters to be found from the considered
sample for the $n$-piece-wise IMF is $2n-1$ when
$m_1$ and $m_{n+1}$ are the minimum  and maximum  of the masses
of the sample.
In the case of $n=4$, which fits also the
region of brown dwarfs (BD), see
\cite{Zaninetti2013a},
the number of  parameters is seven.
In the field of statistical distributions,
the PDF
is usually defined by two parameters.
Examples of two-parameter PDFs are:
the beta, gamma, normal, and lognormal  distributions,
see \cite{evans}.
The lognormal distribution is widely used
to model the IMF  for the
stars, see \cite{Larson1973,Miller1979,Zinnecker1984,Chabrier2003}.
The lognormal distribution is defined in the range
of  $\mathcal{M}\, \in (0, \infty)$  where $\mathcal{M}$
is the mass of the star.
Nevertheless, the stars have minimum and maximum  values.
In an example  from the MAIN SEQUENCE,  an M8 star has
$\mathcal{M} = 0.06 \sunmass$ and an O3 star has
$\mathcal{M} = 120  \sunmass$, see \cite{Cox}.
The presence  of boundaries for the stars
makes the analysis of the
truncated  lognormal, see \cite{Zaninetti2017d},
and of the truncated Lindley PDF attractive.
In the case of the truncated Lindley PDF,
the  analysis of the samples representative of the IMF for
stars
is limited to those  that produce
both parameters $b$ and $c$ positive,
and are therefore suitable
for a fit with the  truncated Lindley distribution.
The statistical parameters are the same of
\cite{Zaninetti2017d}
and are
the merit function $\chi^2$,
the  reduced  merit function $\chi_{red}^2$,
the Akaike information criterion
(AIC),
the number of degrees  of freedom
$NF=n-k$ where
$n$ is the number of bins and $k$ is the number of parameters,
the goodness  of the fit  expressed by
the probability $Q$,
the maximum  distance, $D$, between the theoretical
and the astronomical  DF
and  the  significance  level,  $P_{KS}$,
for the Kolmogorov--Smirnov test (K--S).

\begin{figure*}
\begin{center}
\includegraphics[width=10cm,angle=-90]{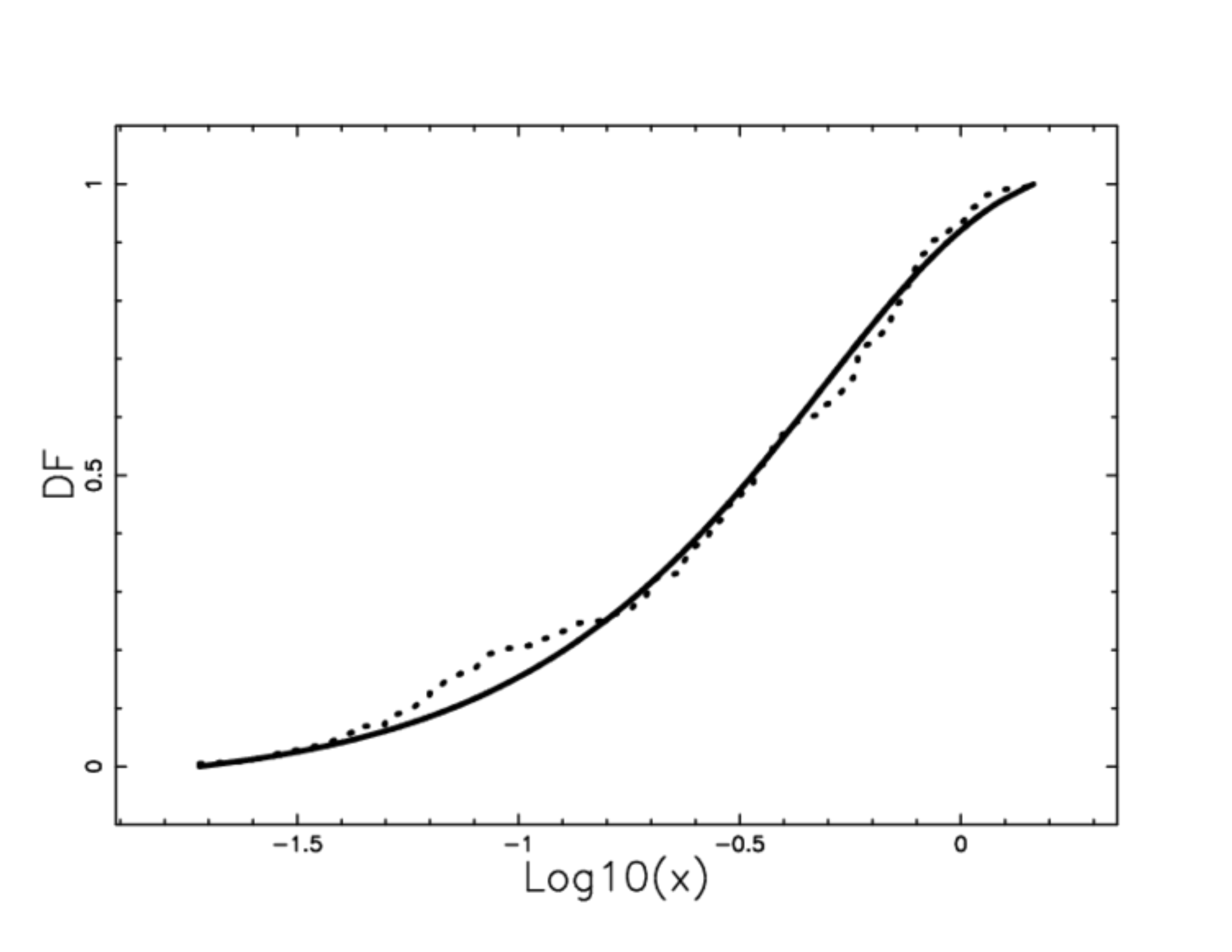}
\end{center}
\caption
{
Empirical DF   of  mass distribution
for    NGC 6611 cluster data
(207 stars + BDs)
when the number of bins, $n$, is 20
(dotted points )
with a superposition of the   truncated  Lindley
DF  (full  line).
Theoretical parameters as in Table
\ref{chi2valuesngc6611}, MLE method.
The  horizontal axis  has a  logarithmic scale.
}
\label{lindley_df_ngc6611}
\end{figure*}

To give an example,  Figure \ref{lindley_df_ngc6611}
reports the   truncated Lindley DF for   NGC 6611
with statistical parameters as in
Table \ref{chi2valuesngc6611}.

\begin{table}[ht!]
\caption
{
Statistical parameters of
NGC 6611  (207 stars + BDs)
in the case of the   truncated Lindley distribution.
The  number of  linear   bins, $n$, is 20.
}
\label{chi2valuesngc6611}
\begin{center}
\resizebox{16cm}{!}
{
\begin{tabular}{|c|c|c|c|c|c|c|c|}
\hline
PDF       & Method &  parameters  &  AIC  & $\chi_{red}^2$
& $Q$  &  D &   $P_{KS}$  \\
\hline
  truncated Lindley & MLE & $b$=0.666, $c= 1.938$,  $x_l$=0.0189, $x_u$=1.46
& 47.75   &  2.48      & $8.4\,10^{-4}$      &   0.065 &  0.332 \\
lognormal &  MLE & $\sigma$=1.029, $m$=0.284   &  71.24&
3.73    & $1.3\,10^{-7}$  &  0.09366 &  0.04959 \\
\hline
\hline
truncated lognormal & MLE & $\sigma$=1.499, $m$=0.478, $x_l$=0.0189, $x_u$=1.46   &
50.96   & 2.68      & $2.8\,10^{-4}$      &  0.0654 &  0.372
\\
\hline
\end{tabular}
}
\end{center}
\end{table}

A careful analysis of Table  \ref{chi2valuesngc6611}
allows to conclude that in the case of NGC 6611
the truncated Lindley PDF produces  a better 
fit  in respect to the lognormal and truncated lognormal PDFs.

\begin{figure*}
\begin{center}
\includegraphics[width=6cm]{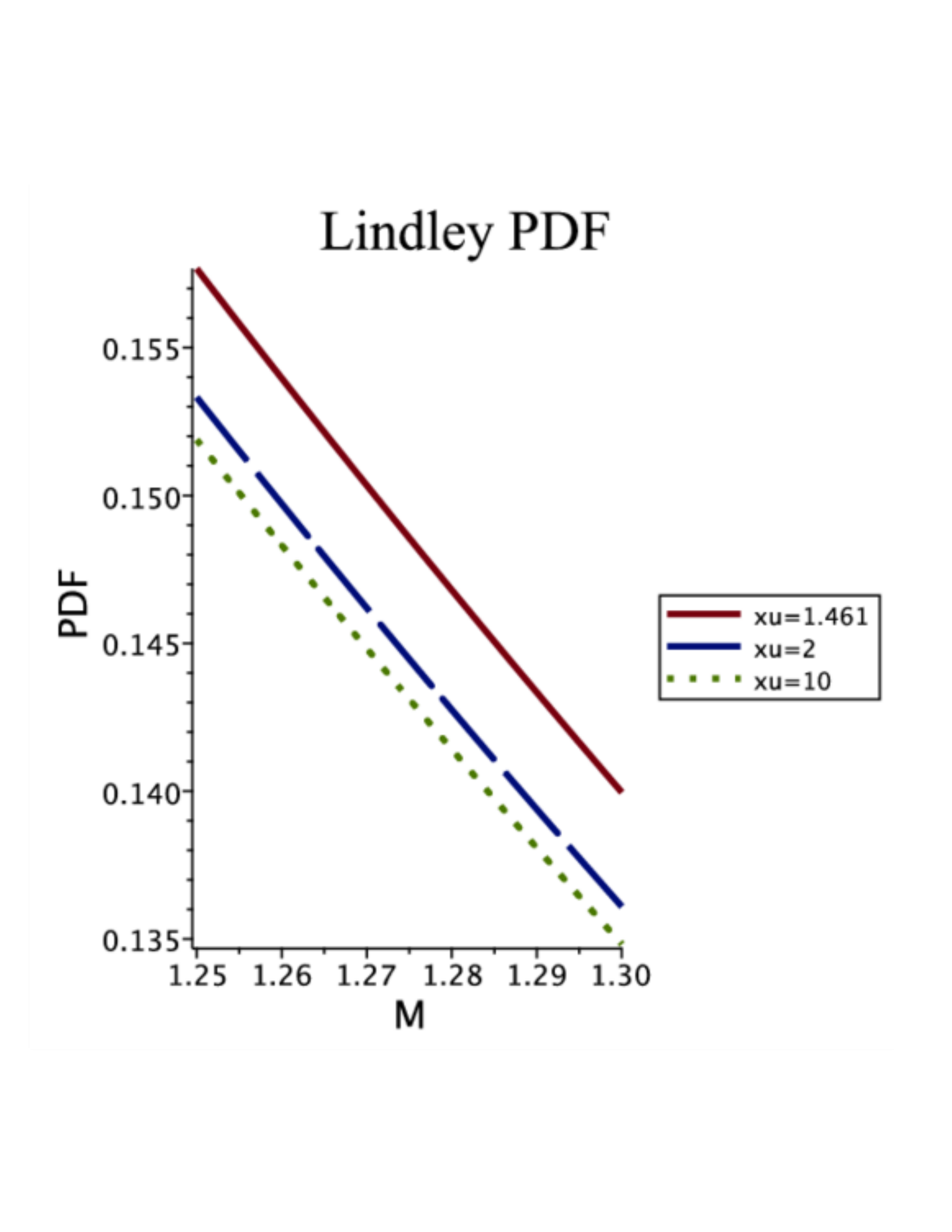}
\end{center}
\caption
{
Double truncated Lindley PDF  with parameters as in Table
\ref{chi2valuesngc6611} and variable $x_u$;
$x_u$ = 1.461 (red  full line  ),
$x_u$ = 2     (blue dashed line)
and
$x_u$ = 10     (green dotted line).
}
\label{lifetime_stars}
\end{figure*}

The lifetime of a star belonging to the MAIN  V, $t_{MS}$,
is
\begin{equation}
\frac{t_{MS}}{t_{\sun}} \approx
\Bigg (
\frac{
\mathcal{M}
}
{
\sunmass
}
\Bigg )^{-2.5}
\quad ,
\end{equation}
where $t_{\sun}$ is the lifetime of the sun, $10^{10}$ yr,
$\mathcal{M}$ is the mass of MAIN V star
and
$\sunmass$ the solar mass,
see \url{http://astronomy.swin.edu.au/cosmos/} for more details.
Figure \ref{lifetime_stars}
reports the modifications of the Lindley PDF
with an increasing upper boundary.
Meanwhile, Table \ref{timestar} reports the correspondence
between  the selected mass and the connected lifetime.

\begin{table}[ht!]
\caption
{
Lifetime of MAIN V star
}
\label{timestar}
\begin{center}
\begin{tabular}{|c|c|}
\hline
Mass in solar units       & lifetime (yr)  \\
\hline
1.461  &  $3.87\times 10^9$               \\
2      &  $1.76\times 10^9$               \\
10     &  $3.16\times 10^7$               \\
\hline
\end{tabular}
\end{center}
\end{table}
For example, in the case  of the cluster NGC 6611,  the upper
limit in mass   will decrease from $1.4\, {M}_{\sun}$
to $1\, {M}_{\sun}$ in $9.9\,10^9~yr$
and after that time the total number of stars
will be the $92.25\%$ of the original number of stars.
The above model allows to see how the time 
modifies  the Hertzsprung-Russell (H-R) diagram,
i.e. the $M_{\mathrm V}$  against $(B-V)$,
in the young clusters of stars.

\section{The luminosity function for galaxies}

\label{lindleylfsec}
In this section, we  review the standard  luminosity function (LF) for
galaxies, we  introduce  a Lindley LF and
a   truncated Lindley LF,
we  then outline the formulae of the photometric maximum
and we  parametrize  the averaged absolute magnitude
as function of the redshift.

\subsection{The  Schechter function }

\label{secshechterlf}
The  Schechter function,
introduced by
\cite{schechter},
provides a useful fit  for the
LF  of galaxies
\begin{equation}
\Phi (L;\alpha,L^*,\Phi^*) dL  = (\frac {\Phi^*}{L^*}) (\frac {L}{L^*})^{\alpha}
\exp \bigl ( {-  \frac {L}{L^*}} \bigr ) dL \quad  ,
\label{equation_schechter}
\end {equation}
here $\alpha$ sets the slope for low values
of $L$, $L^*$ is the
characteristic luminosity and $\Phi^*$ is the normalization.
The equivalent distribution in absolute magnitude is
\begin{equation}
\Phi (M)dM=0.921 \Phi^* 10^{0.4(\alpha +1 ) (M^*-M)}
\exp \bigl ({- 10^{0.4(M^*-M)}} \bigr)  dM \, ,
\label{lfstandard}
\end {equation}
where $M^*$ is the characteristic magnitude as derived from the
data.
We now  introduce  the parameter $h$
which is $H_0/100$, where $H_0$ is the Hubble constant.
The scaling with  $h$ is  $M^* - 5\log_{10}h$ and
$\Phi^* ~h^3~[Mpc^{-3}]$.
The numerical exploration  of  a new LF for galaxies
requires that the $\chi^2_{red}$
is smaller or approximately equal to that
of the  Schechter LF.
As an example,  the LF given by the
generalized gamma distribution with four parameters
gives $\chi^2_{red}$ smaller than that of
the  Schechter LF in the five bands of the Sloan Digital Sky Survey (SDSS)
 galaxies,
see equation (21) an Table II in \cite{Zaninetti2010f}

\subsection{The Lindley LF}

\begin{figure*}
\begin{center}
\includegraphics[width=10cm,angle=-90]{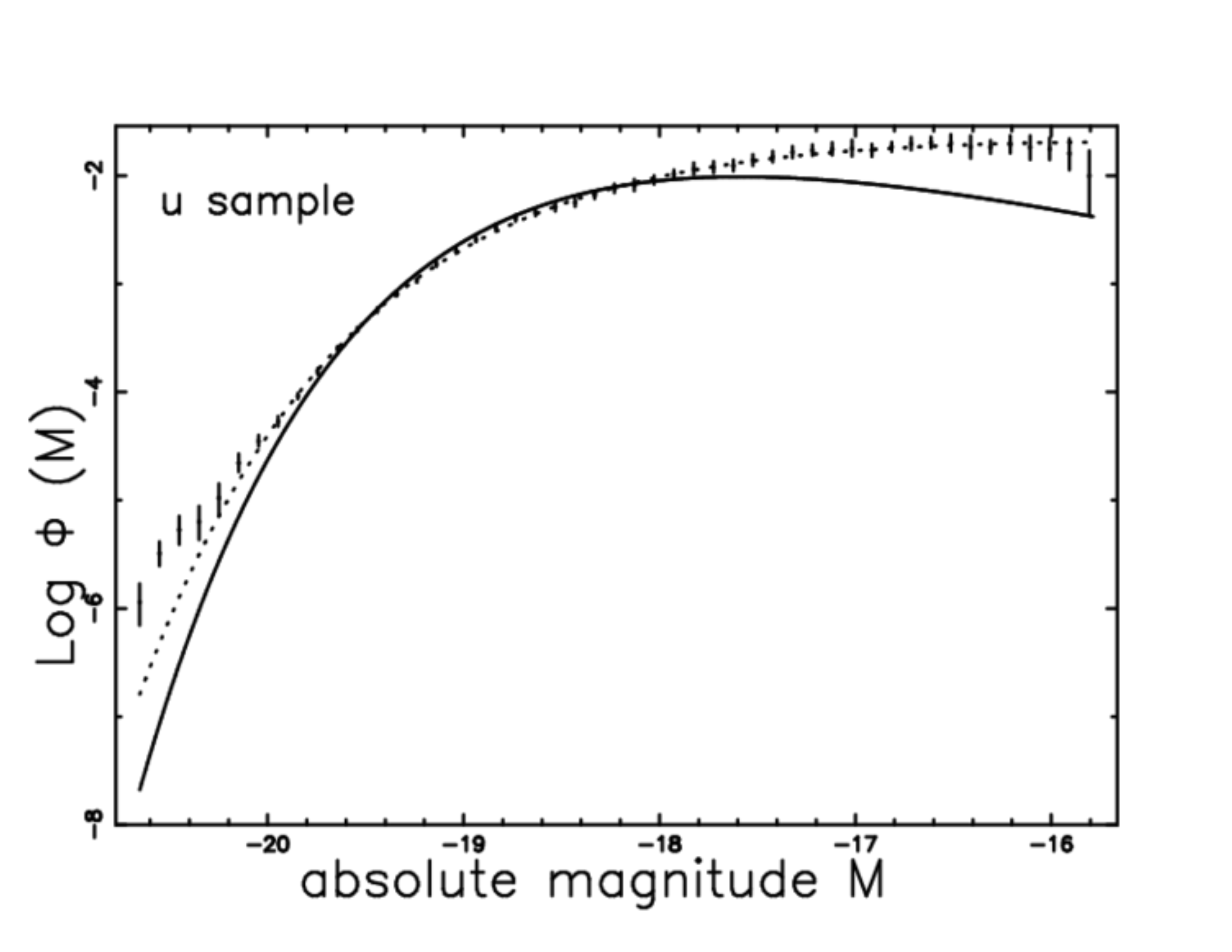}
\end{center}
\caption
{
The luminosity function data of
SDSS($u^*$) are represented with error bars.
The continuous line fit represents the Lindley LF
(\ref{lindleymag})
and the dotted
line represents the Schechter function.
}
\label{due_u}
\end{figure*}

We start with the Lindley PDF with scaling as given by
equation (\ref{lindleypdfscaling}),
\begin{equation}
\Psi(L;c,L^*,\Psi^*) dL  =
\frac
{
{ \Psi^* }\,{c}^{2}{{\rm e}^{-{\frac {cL}{{\it L^*}}}}} \left( L+{
\it L^*} \right)
}
{
{{\it L^*}}^{2} \left( 1+c \right)  dL
}\, dM   \quad ,
\end{equation}
where
$L$  is the luminosity, $L^*$ is the
characteristic luminosity and $\Psi^*$ is the normalization.
The magnitude version is
\begin{equation}
\Psi(M;c,M^*,\Psi^*) dM  =
\frac
{
 0.4\,{ \Psi^* }\,{c}^{2}\ln  \left( 10 \right) {{\rm e}^{-c{10}^{-
 0.4\,M+ 0.4\,{M}^{{\it *}}}}} \left( {10}^{- 0.4\,M+ 0.4\,{M}^{{
\it *}}}+{10}^{- 0.8\,M+ 0.8\,{M}^{{\it *}}} \right)
}
{
1+c
}
\quad  ,
\label{lindleymag}
\end{equation}
where $M$ is the absolute magnitude,
$M^*$   the characteristic magnitude
and $\Psi^*$ is the normalization.
A test is  performed on the $u^*$
band  of SDSS as
 in \cite{Blanton_2003} with data available
at \url{https://cosmo.nyu.edu/blanton/lf.html}.
The Schechter function, the new Lindley LF
represented by formula~(\ref{lindleymag})
and the data are
reported in
Figure~\ref{due_u}, parameters as
in Table \ref{chi2valuelf}.

\begin{table}[ht!]
\caption {
Numerical values values and  $\chi^2_{red}$ of the LFs
applied to  SDSS Galaxies
in the $u^*$ band
 }
\label{chi2valuelf}
\begin{center}
\begin{tabular}{|c|c|c|}
\hline
LF        &   parameters    & $\chi^2_{red}$ \\
\hline
Schechter &
$M^*$= -17.92 ,\, $\alpha$=-0.9,\, $\Phi^* = 0.0114 /Mpc^3$
 & 0.689
  \\
Lindley &  $M^*$= -23.40 ,\, c=214.1 , \, $\Psi^* = 0.0289 /Mpc^3$
&  6.6741   \\
truncated Lindley &  $M^*$= -23.458 ; c= 224.47  
; $\Psi^* =  0.0239 /Mpc^3$ ;$M_l$=-20.653 ; $M_u$=-15.785  
&  6.6739  \\
\hline
\end{tabular}
\end{center}
\end{table}

\subsection{The   truncated Lindley LF}

We start with the   truncated Lindley PDF with scaling as given by
equation (\ref{pdflindleytruncated})
\begin{equation}
\Psi(L;c,L^*,\Psi^*,L_l,L_u) dL  =
\frac
{
{\it \Psi^*}\,{{\rm e}^{-{\frac {cL}{{\it L^*}}}}} \left( L+{\it
L^*} \right) {c}^{2}
}
{
DT
}
\quad ,
\label{lindleylf}
\end{equation}
with
\begin{eqnarray}
DT=
{\it L^*}\, \left( {{\rm e}^{-{\frac {cL_{{l}}}{{\it L^*}}}}}{\it
L^*}\,c+{{\rm e}^{-{\frac {cL_{{l}}}{{\it L^*}}}}}cL_{{l}}-{
{\rm e}^{-{\frac {cL_{{u}}}{{\it L^*}}}}}{\it L^*}\,c-{{\rm e}^{-{
\frac {cL_{{u}}}{{\it L^*}}}}}cL_{{u}}+{{\rm e}^{-{\frac {cL_{{l}}}{
{\it L^*}}}}}{\it L^*}-{{\rm e}^{-{\frac {cL_{{u}}}{{\it L^*}}}}
}{\it L^*} \right) \quad ,
\end{eqnarray}
where
$L$  is the luminosity,  $L^*$ is the
characteristic luminosity,
$L_l$ is the lower boundary in luminosity,
$L_u$ is the upper boundary in luminosity,
 and $\Psi^*$ is the normalization.
The magnitude version is
\begin{equation}
\Psi(M;c,M^*,\Psi^*,M_l,M_u) dM  =
\frac
{
NM
}
{
DT
}
\end{equation}
where
\begin{eqnarray}
NM =
0.4\,{c}^{2}{\it \Psi^*}\, \left( \ln  \left( 2 \right) +\ln
 \left( 5 \right)  \right) {{\rm e}^{c \left( {10}^{- 0.4\,M_{{l}}+
 0.4\,{\it M^*}}+{10}^{ 0.4\,{\it M^*}- 0.4\,M_{{u}}}-{10}^{ 0.4\,
{\it M^*}- 0.4\,M} \right) }}
\times  \nonumber \\
\left( {10}^{ 0.4\,M_{{l}}+ 0.4\,M_{{u
}}}{100}^{ 0.4\,{\it M^*}- 0.4\,M}+{10}^{ 0.4\,M_{{l}}+ 0.4\,{\it
M^*}+ 0.4\,M_{{u}}- 0.4\,M} \right)
\end{eqnarray}
\begin{eqnarray}
DT =
{10}^{ 0.4\,M_{{l}}+ 0.4\,{\it M^*}}{{\rm e}^{c{10}^{- 0.4\,M_{{l}}+
 0.4\,{\it M^*}}}}c+{{\rm e}^{c{10}^{- 0.4\,M_{{l}}+ 0.4\,{\it M^*
}}}}{10}^{ 0.4\,M_{{l}}+ 0.4\,M_{{u}}}c
\nonumber \\
-{10}^{ 0.4\,{\it M^*}+ 0.4\,
M_{{u}}}{{\rm e}^{c{10}^{ 0.4\,{\it M^*}- 0.4\,M_{{u}}}}}c-{{\rm e}^
{c{10}^{ 0.4\,{\it M^*}- 0.4\,M_{{u}}}}}{10}^{ 0.4\,M_{{l}}+ 0.4\,M_
{{u}}}c
\nonumber  \\
+{{\rm e}^{c{10}^{- 0.4\,M_{{l}}+ 0.4\,{\it M^*}}}}{10}^{ 0.4
\,M_{{l}}+ 0.4\,M_{{u}}}-{{\rm e}^{c{10}^{ 0.4\,{\it M^*}- 0.4\,M_{{
u}}}}}{10}^{ 0.4\,M_{{l}}+ 0.4\,M_{{u}}}
\quad ,
\end{eqnarray}
where $M$ is the absolute magnitude,
$M^*$   the characteristic magnitude,
$M_l$   the lower boundary in  magnitude,
$M_u$   the upper boundary in  magnitude and
 $\Psi^*$ is the normalization.
The mean theoretical absolute  magnitude, ${ \langle M \rangle }$,
can  be evaluated as
\begin{equation}
 { \langle M \rangle }
=
\frac
{
\int_{M_l}^{M_u} M \times \Psi(M;c,M^*,\Psi^*,M_l,M_u) dM
}
{
\int_{M_l}^{M_u}  \Psi(M;c,M^*,\Psi^*,M_l,M_u) dM
}
\quad .
\label{meanabsolutetruncated}
\end{equation}
At the moment of writing, the analytical solution
does not exists and the integration should be done
numerically.
Table \ref{chi2valuelf} reports  the parameters 
of the truncated Lindley LF 
from which is possible to conclude that 
the effect of truncation  in the Lindley LF produces a minimum
decrease in the $\chi_{red}^2$ : Lindley  LF with truncation 
$\chi_{red}^2=6.6739$ and    Lindley  LF $\chi_{red}^2=6.6741$.

\subsection{The photometric maximum}

In the pseudo-Euclidean universe,
the  correlation
between expansion velocity  and distance is
\begin {equation}
V= H_0 D  = c_l \, z
\quad  ,
\label {clz}
\end{equation}
where $H_0$ is the Hubble constant, after \cite{Hubble1929},
$H_0 = 100 h \mathrm{\ km\ s}^{-1}\mathrm{\ Mpc}^{-1}$, with $h=1$
when  $h$ is not specified,
$D$ is the distance in $Mpc$,
$c_l$ is  the  light velocity  and
$z$
is the redshift.
In the
pseudo-Euclidean
universe
the flux of radiation,
$ f$,  expressed in $ \frac {L_{\sun}}{Mpc^2}$ units,
where $L_{\sun}$ represents the luminosity of the sun,  is
\begin{equation}
f  = \frac{L}{4 \pi D^2}
\quad ,
\end{equation}
where $D$   represents the distance of the galaxy
expressed in $Mpc$,
and
\begin{equation}
D=\frac{c_l z}{H_0}
\quad  .
\end{equation}
The joint distribution in {\it z}
and {\it f}  for the Schechter LF,
see formula~(1.104) in
 \cite{pad}
or formula~(1.117)
in
\cite{Padmanabhan_III_2002},
 is
\begin{equation}
\frac{dN}{d\Omega dz df} =
4 \pi  \bigl ( \frac {c_l}{H_0} \bigr )^5    z^4 \Phi (\frac{z^2}{z_{crit}^2})
\label{nfunctionzschechter}
\quad ,
\end {equation}
where $d\Omega$, $dz$ and  $ df $ represent
the differential of
the solid angle,
the redshift and the flux respectively
and     $\Phi$ is the Schechter LF.
The critical value of $z$,   $z_{crit}$, is
\begin{equation}
 z_{crit}^2 = \frac {H_0^2  L^* } {4 \pi f c_l^2}
\quad ,
\end{equation}
where  $L^*$ has been defined  in Section \ref{secshechterlf}.
The number of galaxies in $z$  and $f$ for the
Schechter LF
as given by
formula~(\ref{nfunctionzschechter})
has a maximum  at  $z=z_{pos-max}$,
where
\begin{equation}
 z_{pos-max} = z_{crit}  \sqrt {\alpha +2 }
\quad ,
\end{equation}
which  can be re-expressed   as
\begin{equation}
 z_{pos-max}(f) =
\frac
{
\sqrt {2+\alpha}\sqrt {{10}^{ 0.4\,{\it M_{\sun}}- 0.4\,{\it M^*}}}{
\it H_0}
}
{
2\,\sqrt {\pi }\sqrt {f}{\it c_l}
}
\quad  ,
\label{zmax_sch}
\end{equation}
where $M_{\sun}$ is the reference magnitude
of the sun at the considered bandpass.
The position of the maximum in redshift for the Schechter LF  depends
from the flux of the selected astronomical band, $f$,
and from the two parameter which characterizes 
the Schechter LF: $\alpha$ and  $M^*$.

More details can be found  in \cite{Zaninetti2014b}.

The joint distribution in
$z$
and $f$
for galaxies  for  the
Lindley LF, see equation (\ref{lindleylf}),
is
\begin{equation}
\frac{dN}{d\Omega dz df} =
\frac
{
4\,{z}^{4}{c}^{2}{{\rm e}^{-{\frac {c{z}^{2}}{{z_{{{\it crit}}}}^{2}}}
}}{c_{{l}}}^{5}\pi\, \left( {z}^{2}+{z_{{{\it crit}}}}^{2} \right)
}
{
\left( 1+c \right) {H_{{0}}}^{5}{\it L^*}\,{z_{{{\it crit}}}}^{2}
}
\quad .
\label{nfunctionzlindley}
\end{equation}
The maximum in the number of galaxies for the Lindley LF
as function of $z_{crit}$ is at
\begin{equation}
 z_{pos-max}(z_{crit}) =
\frac
{
\sqrt {2}\sqrt {-c+3+\sqrt {{c}^{2}+2\,c+9}}z_{{{\it crit}}}
}
{
2\,\sqrt {c}
}
\quad ,
\end{equation}
or  as  function of the flux $f$
\begin{equation}
 z_{pos-max}(f) =
\frac
{
\sqrt {2}\sqrt {-c+3+\sqrt {{c}^{2}+2\,c+9}}\sqrt {{10}^{ 0.4\,{\it
M_{\sun}}- 0.4\,{\it M^*}}}H_{{0}}
}
{
4\,\sqrt {c}\sqrt {\pi}\sqrt {f}c_{{l}}
}
\quad ,
\end{equation}
or  as a function of the apparent magnitude $m$
\begin{equation}
 z_{pos-max}(m) =
\frac
{
5\,10^{-6}\,\sqrt {2}\sqrt {-c+3+\sqrt {{c}^{2}+2\,c+9}}\sqrt
{{10}^{ 0.4\,{\it M_{\sun}}- 0.4\,{\it M^*}}}H_{{0}}
}
{
\sqrt {c}\sqrt {{{\rm e}^{ 0.921034\,{\it M_{\sun}}- 0.921034\,m}}
}c_{{l}}
}
\quad  .
\end{equation}
The position of the maximum in redshift for the Lindley  LF  depends
from the flux of the selected astronomical band, $f$,
or the selected apparent magnitude, $m$,
and from the two parameter which characterizes 
the Lindley LF: $c$ and  $M^*$.

The mean redshift   for galaxies  $ { \langle z \rangle } $
can be defined as
\begin{equation}
 { \langle z \rangle }
=
\frac
{
\int_0^\infty  z \, \frac{dN}{d\Omega dz df} dz
}
{
\int_0^\infty  \frac{dN}{d\Omega dz df}  dz
}
\quad .
\end{equation}
The mean redshift
for the Lindley LF
as  function of $z_{crit}$
is
\begin{equation}
 { \langle z \rangle } (z_{crit}) =
\frac
{
16\,z_{{{\it crit}}} \left( c+3 \right)
}
{
3\,\sqrt {\pi}\sqrt {c} \left( 2\,c+5 \right)
}
\quad ,
\end{equation}
or  as a function of the flux
\begin{equation}
 { \langle z \rangle } (f) =
\frac
{
8\,\sqrt {\pi\,f{10}^{ 0.4\,{\it M_{\sun}}- 0.4\,{\it M^*}}}H_{{0}}
 \left( c+3 \right)
}
{
3\,{\pi}^{3/2}fc_{{l}}\sqrt {c} \left( 2\,c+5 \right)
}
\end{equation}
or  as a function of the apparent magnitude
\begin{equation}
 { \langle z \rangle } (m) =
\frac
{
 3.009\,10^{-5}\,\sqrt {{{\rm e}^{ 0.921034\,{\it M_{\sun}}-
 0.921034\,m}}{10}^{ 0.4\,{\it M_{\sun}}- 0.4\,{\it M^*}}}H_{{0}}
 \left( c+3 \right)
}
{
{{\rm e}^{ 0.9210340374\,{\it M_{\sun}}- 0.9210340374\,m}}c_{{l}}\sqrt {c}
 \left( 2\,c+5 \right)
}
\quad .
\end{equation}

\begin{figure}
\begin{center}
\includegraphics[width=6cm,angle=-90]{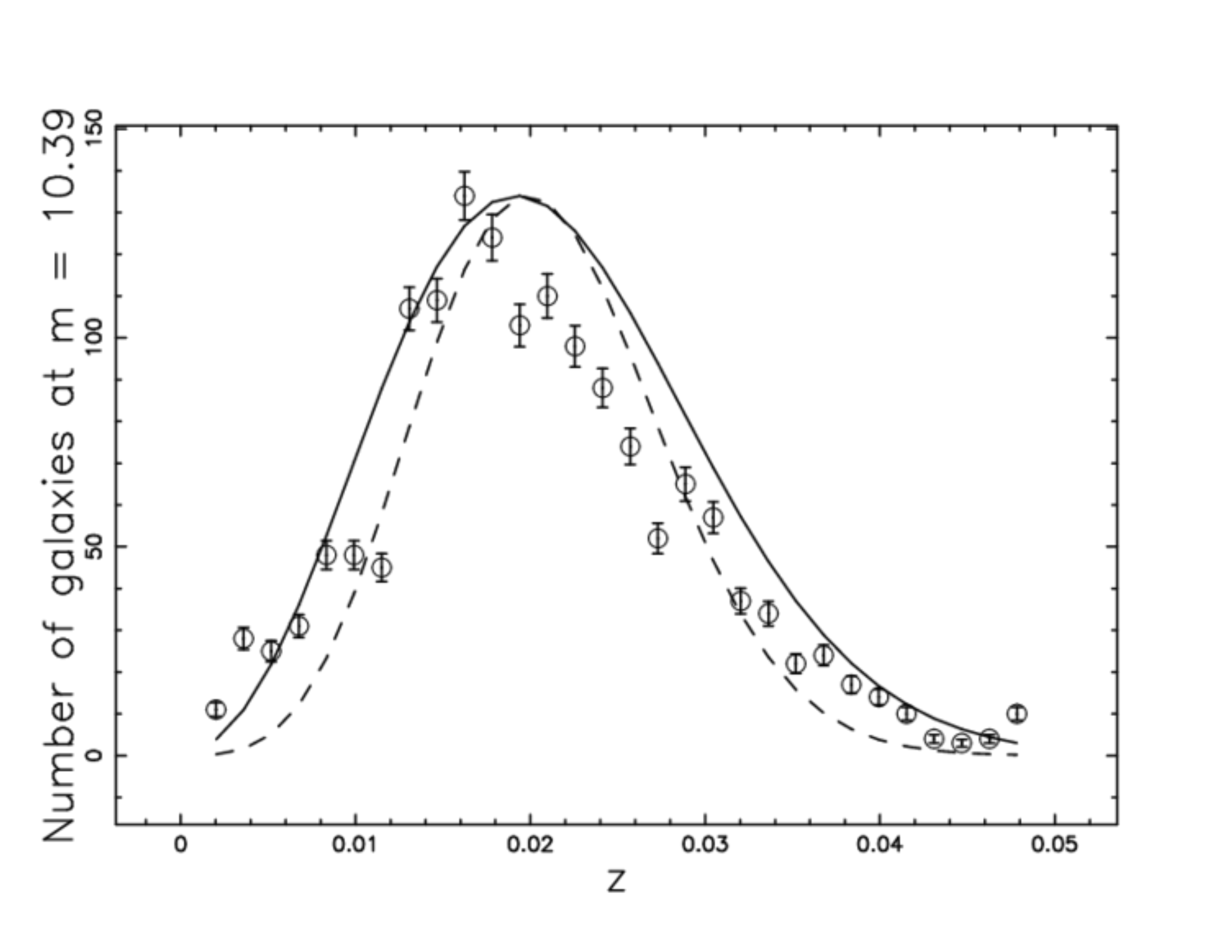}
\end {center}
\caption{
The galaxies  of the 2MRS with
$ 10.28    \leq  m   \leq 10.44  $  or
$ 1202409  \frac {L_{\sun}}{Mpc^2} \leq
f \leq  1384350 \frac {L_{\sun}}{Mpc^2}$
are  organized in frequencies versus
heliocentric  redshift,
(empty circles);
the error bar is given by the square root of the frequency.
The maximum frequency of observed galaxies is
at  $z=0.018$.
The full line is the theoretical curve
generated by
$\frac{dN}{d\Omega dz df}(z)$
as given by the application of the Schechter LF
which  is equation~(\ref{nfunctionzschechter})
and the dashed line
represents the Lindley  LF
which  is equation~(\ref{nfunctionzlindley}).
The parameters are  the same of Table \ref{chi2valuemax},
$\chi^2= 198$  for the Schechter LF    and
$\chi^2= 191$  for the Lindley   LF.
}
          \label{lindley_photo_max}%
    \end{figure}

Figure \ref{lindley_photo_max}
reports the number of  observed  galaxies
of the  2MASS Redshift Survey (2MRS)  catalog for  a given
apparent magnitude  and
the two theoretical curves are analyzed
with same parameters as in Table \ref{chi2valuemax}.
These parameters are derived in  such a way
that the $\chi^2$ is minimum.
Therefore, this is a new method to derive the parameters which
characterize the two LFs here adopted
without using the samples for  the LF such
as the  five bands of SDSS galaxies.

\begin{table}[ht!]
\caption {
Numerical values values and  $\chi^2_{red}$ of the two LFs
applied to $K_S$ band
(2MASS Kron magnitudes)
when  $M_{\sun}$=3.39 \,. }
\label{chi2valuemax}
\begin{center}
\begin{tabular}{|c|c|c|}
\hline
LF        &   parameters    & $\chi^2_{red}$ \\
\hline
 Schechter  &
$M^*$= -23.289, $\alpha$=-0.794,
$\Phi^* = 0.0128 /Mpc^3$
 &
 7.08 \\
 \hline
Lindley &
$M^*$=-23.7  , $c$= 2.8, $\Phi^* =   0.0289 /Mpc^3$ &
 6.84 \\
\hline
\end{tabular}
\end{center}
\end{table}

\subsection{Averaged absolute magnitude}

\begin{figure*}
\begin{center}
\includegraphics[width=6cm,angle=-90]{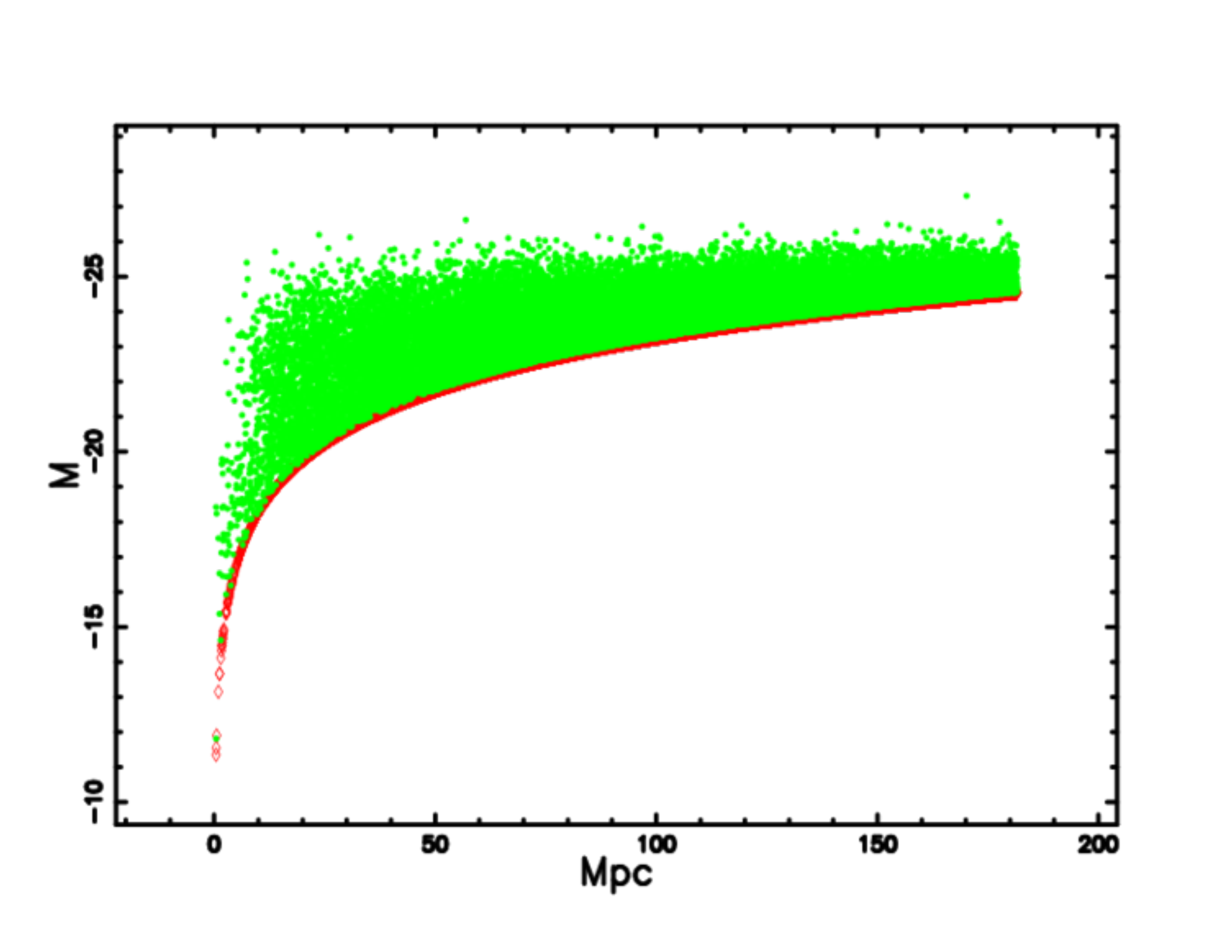}
\end {center}
\caption{
The absolute magnitude $M$ of
36,474 galaxies belonging to the 2MRS
when $\mathcal{M_{\sun}}$ = 3.39 and
$H_{0}=70 \mathrm{\ km\ s}^{-1}\mathrm{\ Mpc}^{-1}$
(green points).
The lower theoretical curve as represented by
Equation~(\ref{absolutel}) is shown as the
red-thick line when $m_L$=11.75.
}
 \label{bias_2mrs}%
 \end{figure*}

\begin{figure*}
\begin{center}
\includegraphics[width=6cm,angle=-90]{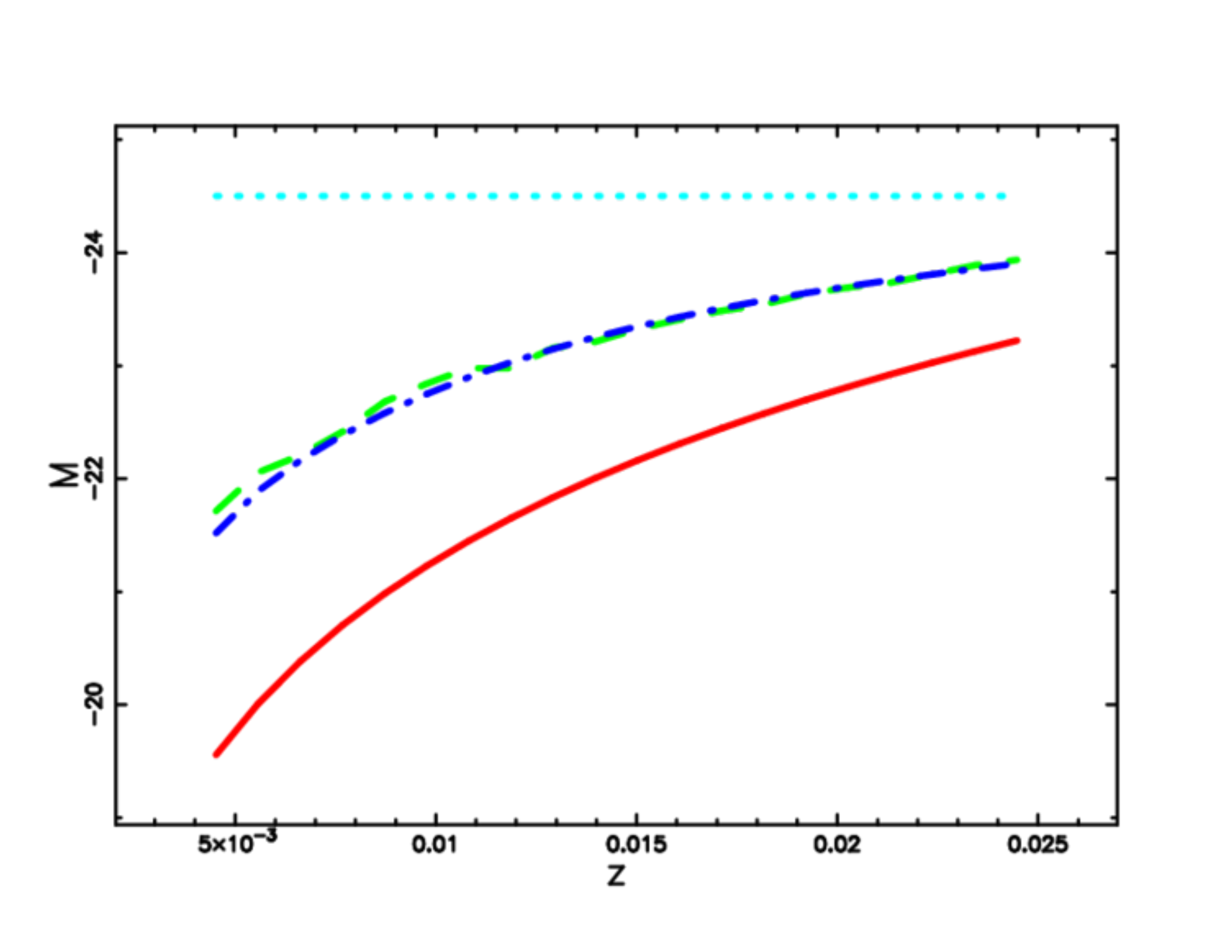}
\end {center}
\caption
{
Averaged absolute magnitude of the
galaxies belonging to the 2MRS
(green-dashed line),
theoretical averaged absolute magnitude
for the   truncated Lindley  LF
(blue dash-dot-dash-dot line)
as given by equation (\ref{meanabsolutetruncated}),
lower theoretical curve as represented by
Equation~(\ref{absolutel})
(red line)
and
minimum absolute magnitude observed (cyan dotted line).
}
 \label{bias_2mrs_theo}%
 \end{figure*}

We now introduce the concept of
limiting apparent magnitude.
The observable absolute magnitude as a function of the
limiting apparent magnitude, $m_L$, is
\begin{equation}
M_L =
m_{{L}}-5\,{\it \log_{10}} \left( {\frac {{\it c}\,z}{H_{{0}}}}
 \right) -25
\quad .
\label{absolutel}
\end{equation}
Figure \ref{bias_2mrs} presents such a curve
and the galaxies of the 2MRS.
We now compare the  theoretical averaged  absolute magnitude
of  the   truncated Lindley LF,
see  equation (\ref{meanabsolutetruncated}),
with the observed averaged absolute magnitude of 2MRS
as function of the redshift.
To fit the  data
we assumed the following empirical  dependence with redshift
for the characteristic   magnitude of the
  truncated Lindley LF
\begin{equation}
M^* = -25.14 + 3\Big (1- \sqrt{ \frac{z-z_{min}} {z_{max}-z_{min}}} ~\Big)
\quad  .
\label{mstarz}
\end{equation}
This relationship models the decrease of the characteristic
absolute magnitude as function of the redshift
and allows us to match observational and theoretical data.
The lower bound  in absolute magnitude  is given
by the minimum magnitude of the selected bin,
the upper bound  is given by
equation (\ref {absolutel}),
the characteristic magnitude varies according
to  equation (\ref {mstarz})
and  Figure \ref{bias_2mrs_theo} reports
the comparison between theoretical and observed absolute magnitude
for 2MRS.

\section{Conclusions}

{\bf Statistical Distributions}
We introduced the Lindley distribution with
scale and the
truncated Lindley distribution.
The parameters of the  Lindley distribution with scale
can be found with the method of the matching moments.
In the the case of
the truncated
Lindley distribution
the MLE  is used  to estimate
the unknown parameters.

{\bf Application to the stars}
To fit the IMF for stars
with the   truncated Lindley PDF,
the
 parameters $b$ and $c$, which is deduced from the astronomical  sample,
should be positive.
This is the case of NGC 6611  (207 stars + BDs),
for which the
reduced  merit function
is smaller for the   truncated Lindley distribution
in respect to the lognormal and truncated lognormal distribution,
see Table \ref{chi2valuesngc6611}.

{\bf Application to the galaxies}
The Lindley LF for galaxies is characterized
by   a higher   reduced  merit function
in respect to  the  Schechter  LF for the case of
SDSS Galaxies
in the $u^*$ band, see Table \ref{chi2valuelf}.
Conversely  the Lindley LF for galaxies
produces a lower  value  of
the merit function when the
photometric maximum of 2MRS is modeled
in respect to  the  Schechter  model for the maximum,
see Figure \ref{lindley_photo_max}.
The truncated Lindley LF produces an acceptable
model  for the averaged absolute magnitude of the
galaxies belonging to the 2MRS, 
see Figure \ref{bias_2mrs_theo}.

\appendix
\setcounter{equation}{0}
\renewcommand{\theequation}{\thesection.\arabic{equation}}

\section{Other parameters of the Lindley distribution}
\label{appendixdl}

The rth moment about the origin for the Lindley distribution is,
$\mu^{\prime}_r$,
is
\begin{equation}
\mu^{\prime}_r =
\frac
{
{c}^{-r}\Gamma \left( r+2 \right) +{c}^{1-r}\Gamma \left( r+1 \right)
}
{
1+c
}
\quad ,
\end{equation}
where
\begin{equation}
\mathop{\Gamma\/}\nolimits\!\left(z\right)
=\int_{0}^{\infty}e^{{-t}}t^{{z-1}}dt
\quad ,
\end{equation}
is the gamma function, see \cite{NIST2010}.
The central moments, $\mu_r$, are
\begin{subequations}
\begin{align}
\mu_3& =
\frac
{
2\,{c}^{3}+12\,{c}^{2}+12\,c+4
}
{
{c}^{3} \left( 1+c \right) ^{3}
}  \\
\mu_4& =
\frac
{
9\,{c}^{4}+72\,{c}^{3}+132\,{c}^{2}+96\,c+24
}
{
{c}^{4} \left( 1+c \right) ^{4}
}
\end{align}
\end{subequations}

being  $\mu_2=\sigma^2$.
Is impossible to evaluate the median in
a closed form   and
therefore we introduce
an approximated  distribution function, $F_{2,2}$ in terms of
the Pad\'e rational polynomial
approximation,
after \cite{Pade1892},
 of degree 2 in the numerator
and degree 2 in the denominator about the point $x=0$
\begin{equation}
F_{2,2} =
\frac
{
6\,x{c}^{2} \left( 2\,{c}^{2}-cx-4\,c+3\,x+6 \right)
}
{
\left( {c}^{4}{x}^{2}-4\,{c}^{3}{x}^{2}+6\,{c}^{3}x+7\,{c}^{2}{x}^{2}
-18\,{c}^{2}x+12\,{c}^{2}+24\,cx-24\,c+36 \right)  \left( 1+c \right)
}
\end{equation}
The approximated median, $m_{2,2}$ turns out
to be
\begin{equation}
m_{2,2}
=
\frac
{
9\,{c}^{3}-18\,{c}^{2}-\sqrt {69\,{c}^{6}-276\,{c}^{5}+690\,{c}^{4}-
876\,{c}^{3}+1101\,{c}^{2}-984\,c+1188}+33\,c-12
}
{
\left( {c}^{3}-3\,{c}^{2}+15\,c-29 \right) c
}
\quad .
\end{equation}
The percent error, $\delta$, in the evaluation of
the approximated median   is $\delta=1.179\,\%$ at
c=0.5 and  $\delta=0.077\,\%$ at
c=2.

\section{Moments for the Lindley distribution with scale}

\label{appendixldscale}

The rth moment about the origin for the Lindley distribution with scale,
$\mu^{\prime}_{r,s}$,
is
\begin{equation}
\mu^{\prime}_{r,s} =
{\frac {{c}^{-r}{b}^{r}\Gamma \left( r+2 \right) }{1+c}}+{\frac {{c}^{
1-r}{b}^{r}\Gamma \left( r+1 \right) }{1+c}}
\quad .
\end{equation}
The central moments, $\mu_{r,s}$, are
\begin{subequations}
\begin{align}
\mu_{3,s}& =
\frac
{
2\,{b}^{3} \left( {c}^{3}+6\,{c}^{2}+6\,c+2 \right)
}
{
{c}^{3} \left( {c}^{3}+3\,{c}^{2}+3\,c+1 \right)
}
\\
\mu_{4,s}& =
\frac
{
3\,{b}^{4} \left( 3\,{c}^{4}+24\,{c}^{3}+44\,{c}^{2}+32\,c+8 \right)
}
{
{c}^{4} \left( {c}^{4}+4\,{c}^{3}+6\,{c}^{2}+4\,c+1 \right)
}
\quad .
\end{align}
\end{subequations}

\section{The parameters of the truncated Lindley distribution}

\label{appendixmle}

The parameters of the truncated Lindley distribution
can be obtained from empirical data by
the
maximum likelihood estimators  (MLE)
and by the evaluation of the minimum and maximum elements
of the sample.
Consider a  sample  ${\mathcal X}=x_1, x_2, \dots , x_n$ and let
$x_{(1)} \geq x_{(2)} \geq \dots \geq x_{(n)}$ denote
their order statistics, so that
$x_{(1)}=\max(x_1, x_2, \dots, x_n)$, $x_{(n)}=\min(x_1, x_2, \dots, x_n)$.
The first two parameters $x_l$ and $x_u$
are
\begin{equation}
{x_l}=x_{(n)}, \qquad { x_u}=x_{(1)}
\quad  .
\label{eq:firstpar}
\end{equation}
The MLE is obtained by maximizing
\begin{equation}
\Lambda = \sum_i^n \ln(f_t(x_i;b,c,x_l,x_u)).
\end{equation}
The two derivatives $\frac{\partial \Lambda}{\partial b} =0$ and
$\frac{\partial \Lambda}{\partial c}=0 $  generate two
non-linear equations in
 $b$ and $c$ which can be solved numerically,
 we used FORTRAN subroutine SNSQE in \cite{Kahaner1989},
\begin{equation}
\frac{\partial \Lambda}{\partial b}=
\frac
{PNB}
{
{b}^{2} \left( -{{\rm e}^{-{\frac {cx_{{u}}}{b}}}}bc-{{\rm e}^{-{
\frac {cx_{{u}}}{b}}}}cx_{{u}}+{{\rm e}^{-{\frac {cx_{{l}}}{b}}}}bc+{
{\rm e}^{-{\frac {cx_{{l}}}{b}}}}cx_{{l}}-{{\rm e}^{-{\frac {cx_{{u}}
}{b}}}}b+{{\rm e}^{-{\frac {cx_{{l}}}{b}}}}b \right)
}
=0
\quad ,
\end{equation}
where
\begin{eqnarray}
PNB=
-{{\rm e}^{-{\frac {cx_{{u}}}{b}}}}b{c}^{2}nx_{{u}}-{{\rm e}^{-{\frac
{cx_{{u}}}{b}}}}{c}^{2}n{x_{{u}}}^{2}+{{\rm e}^{-{\frac {cx_{{l}}}{b}}
}}b{c}^{2}nx_{{l}}+{{\rm e}^{-{\frac {cx_{{l}}}{b}}}}{c}^{2}n{x_{{l}}}
^{2}-2\,{{\rm e}^{-{\frac {cx_{{u}}}{b}}}}{b}^{2}cn
\nonumber \\
-2\,{{\rm e}^{-{
\frac {cx_{{u}}}{b}}}}bcnx_{{u}}+2\,{{\rm e}^{-{\frac {cx_{{l}}}{b}}}}
{b}^{2}cn+2\,{{\rm e}^{-{\frac {cx_{{l}}}{b}}}}bcnx_{{l}}+\sum _{i=1}^
{n}{\frac {cx_{{i}}b+c{x_{{i}}}^{2}+{b}^{2}}{x_{{i}}+b}}{{\rm e}^{-{
\frac {cx_{{u}}}{b}}}}bc
\nonumber \\
+\sum _{i=1}^{n}{\frac {cx_{{i}}b+c{x_{{i}}}^{
2}+{b}^{2}}{x_{{i}}+b}}{{\rm e}^{-{\frac {cx_{{u}}}{b}}}}cx_{{u}}-
\sum _{i=1}^{n}{\frac {cx_{{i}}b+c{x_{{i}}}^{2}+{b}^{2}}{x_{{i}}+b}}{
{\rm e}^{-{\frac {cx_{{l}}}{b}}}}bc
\nonumber\\
-\sum _{i=1}^{n}{\frac {cx_{{i}}b+c
{x_{{i}}}^{2}+{b}^{2}}{x_{{i}}+b}}{{\rm e}^{-{\frac {cx_{{l}}}{b}}}}cx
_{{l}}
-2\,{{\rm e}^{-{\frac {cx_{{u}}}{b}}}}{b}^{2}n+2\,{{\rm e}^{-{
\frac {cx_{{l}}}{b}}}}{b}^{2}n+\sum _{i=1}^{n}{\frac {cx_{{i}}b+c{x_{{
i}}}^{2}+{b}^{2}}{x_{{i}}+b}}{{\rm e}^{-{\frac {cx_{{u}}}{b}}}}b
\nonumber \\
-\sum
_{i=1}^{n}{\frac {cx_{{i}}b+c{x_{{i}}}^{2}+{b}^{2}}{x_{{i}}+b}}{
{\rm e}^{-{\frac {cx_{{l}}}{b}}}}b
\quad .
\end{eqnarray}
and
\begin{equation}
\frac{\partial \Lambda}{\partial c}= \frac{PNC}
{
cb \left( -{{\rm e}^{-{\frac {cx_{{u}}}{b}}}}bc-{{\rm e}^{-{\frac {cx_
{{u}}}{b}}}}cx_{{u}}+{{\rm e}^{-{\frac {cx_{{l}}}{b}}}}bc+{{\rm e}^{-{
\frac {cx_{{l}}}{b}}}}cx_{{l}}-{{\rm e}^{-{\frac {cx_{{u}}}{b}}}}b+{
{\rm e}^{-{\frac {cx_{{l}}}{b}}}}b \right)
}
=0,
\end{equation}
where
\begin{eqnarray}
PNC=
-{{\rm e}^{-{\frac {cx_{{u}}}{b}}}}b{c}^{2}nx_{{u}}-{{\rm e}^{-{\frac
{cx_{{u}}}{b}}}}{c}^{2}n{x_{{u}}}^{2}+{{\rm e}^{-{\frac {cx_{{l}}}{b}}
}}b{c}^{2}nx_{{l}}+{{\rm e}^{-{\frac {cx_{{l}}}{b}}}}{c}^{2}n{x_{{l}}}
^{2}+{{\rm e}^{-{\frac {cx_{{u}}}{b}}}}\sum _{i=1}^{n}x_{{i}}b{c}^{2}
\nonumber  \\
+
{{\rm e}^{-{\frac {cx_{{u}}}{b}}}}\sum _{i=1}^{n}x_{{i}}{c}^{2}x_{{u}}
-{{\rm e}^{-{\frac {cx_{{u}}}{b}}}}{b}^{2}cn-2\,{{\rm e}^{-{\frac {cx_
{{u}}}{b}}}}bcnx_{{u}}-{{\rm e}^{-{\frac {cx_{{l}}}{b}}}}\sum _{i=1}^{
n}x_{{i}}b{c}^{2}-{{\rm e}^{-{\frac {cx_{{l}}}{b}}}}\sum _{i=1}^{n}x_{
{i}}{c}^{2}x_{{l}}
\nonumber \\
+{{\rm e}^{-{\frac {cx_{{l}}}{b}}}}{b}^{2}cn
+2\,{
{\rm e}^{-{\frac {cx_{{l}}}{b}}}}bcnx_{{l}}+{{\rm e}^{-{\frac {cx_{{u}
}}{b}}}}\sum _{i=1}^{n}x_{{i}}bc-2\,{{\rm e}^{-{\frac {cx_{{u}}}{b}}}}
{b}^{2}n
\nonumber \\
-{{\rm e}^{-{\frac {cx_{{l}}}{b}}}}\sum _{i=1}^{n}x_{{i}}bc+2
\,{{\rm e}^{-{\frac {cx_{{l}}}{b}}}}{b}^{2}n
\quad .
\end{eqnarray}
\providecommand{\newblock}{}

\end{document}